\documentclass[a4paper,twoside]{article}

\usepackage{epsfig}
\usepackage{subcaption}
\usepackage{calc}
\usepackage{amssymb}
\usepackage{amstext}
\usepackage{amsmath}
\usepackage{graphicx}
\usepackage{amsthm}
\usepackage{multicol}
\usepackage{fancyhdr}
\usepackage{pslatex}
\usepackage{hyperref}
\usepackage{apalike}
\usepackage{inputenc}
\usepackage{algorithm2e}
\usepackage{xcolor}
\usepackage[bottom]{footmisc}
\usepackage{SCITEPRESS}     
\definecolor{red}{RGB}{255, 0, 0}

\fancypagestyle{firstpage}{
    \fancyhead[C]{%
        \color{red}
        \begin{tabular}{|c|}\hline\color{red} To be presented at the 16th International Conference on Agents and Artificial Intelligence (ICAART 2024) \\ \color{red} in-person in Rome, Italy, 24th - 26th February 2024. \\ \hline\end{tabular}%
    }
    
    \fancyhead[L]{}
    \fancyfoot[C]{}
}

\begin{document}

\thispagestyle{firstpage}
\title{DeepTraderX: Challenging Conventional Trading Strategies with Deep Learning in Multi-Threaded Market Simulations}

\author{\authorname{Armand Mihai Cismaru\sup{1}\orcidAuthor{0009-0007-6374-4639}}
\affiliation{\sup{1}Department of Computer Science, University of Bristol, Woodland Road, Bristol, United Kingdom}
\email{armandcismaru@gmail.com}
}


\abstract{In this paper, we introduce \emph{DeepTraderX} (DTX), a simple Deep Learning-based trader, and present results that demonstrate its performance in a multi-threaded market simulation. In a total of about 500 simulated market days, DTX has learned solely by watching the prices that other strategies produce. By doing this, it has successfully created a mapping from market data to quotes, either bid or ask orders, to place for an asset.
Trained on historical Level-2 market data, i.e., the \emph{Limit Order Book} (LOB) for specific tradable assets, DTX processes the market state $S$ at each timestep $T$ to determine a price $P$ for market orders. The market data used in both training and testing was generated from unique market schedules based on real historic stock market data. DTX was tested extensively against the best strategies in the literature, with its results validated by statistical analysis. Our findings underscore DTX's capability to rival, and in many instances, surpass, the performance of public-domain traders, including those that outclass human traders, emphasising the efficiency of simple models, as this is required to succeed in intricate multi-threaded simulations. This highlights the potential of leveraging "\emph{black-box}" Deep Learning systems to create more efficient financial markets.}

\onecolumn \maketitle \normalsize \setcounter{footnote}{0} \vfill

\section{\uppercase{Introduction}}
\label{sec:introduction}

Recent advancements in computing have catalysed profound transformations in Artificial Intelligence (AI), which now permeates many facets of our daily lives.

One area impacted by this transformation is the financial sector, or more specifically, financial markets. They are made up of traders, whether human or machine, with the core objective of being as profitable as possible. We call "algorithmic traders" the software-driven entities that have replaced human traders, performing based on pre-defined, complex algorithms derived from complex financial engineering. As markets and technology evolve together, the need for adaptability to fluctuating conditions is of foremost importance. Enter the age of AI traders: more efficient, enabled to make decisions based on instantaneous data analysis, and navigating markets better than their predecessors.

However, the true paradigm shift is heralded by the rise of Deep Learning. Its changing potential is evident across sectors, from chatbots to advanced medical diagnostics. Deep Learning Neural Networks (DLNNs), modelled after human neural pathways \cite{Shetty2020Diving}, are at the forefront of this AI revolution. Their applications span diverse domains such as speech recognition, natural language processing, and even cancer detection \cite{Abed2022Deep}. Recent studies underscore the effectiveness of DLNN-based traders, which have demonstrated capabilities rivalling, if not exceeding, traditional algorithmic traders \cite{Calvez2018Deep}. Moreover, the rapid democratisation of computational power has led to increasingly sophisticated market simulations, enabling a vast number of research prospects — especially for the AI community.

Algorithmic traders execute the majority of daily trades in a market, processing millions of transactions at sub-second rates. While much of the existing literature evaluates trading strategies in simplified market simulations, the intricate and asynchronous nature of real-world financial markets often remains unaddressed. The purpose of this work is to bridge this research gap in the literature with these core contributions:
\begin{itemize}
    \item Train an intelligent trader based on an extended and enhanced, proven DLNN architecture.
    \item Adapt an asynchronous market simulator to use our autonomous strategy and enable a solid base for experiments.
    \item Evaluate the performance of our trader against other traders in the literature based on profits obtained.
    \item Validate the results by performing the relevant statistical analysis and discussing the successes and shortcomings of our model.
\end{itemize}

The aim is to build and tune a system that could outperform existing strategies, in that case raising questions such as: Why rely on traditional methods when AI-powered strategies offer superior performance?

A positive result could have an impact in the real world, with the only barrier represented by the access to some of the LOB data the model requires. Customer limit prices are not public when trading, so owning that kind of historical data would prove to be great leverage. In the case of a negative result, this research would prove useful underlying causes and ascertain the vulnerabilities that a DLNN trader has when deployed in a realistic setting, albeit with the caveat of accessing certain proprietary LOB data. Our exploration is both an homage and an extension of the efforts of two previous pieces of research, DeepTrader \cite{Wray2020Automated} and the Threaded Bristol Stock Exchange \cite{rollins2020trading}, seeking to chart new horizons in the confluence of AI and financial trading.


\subsubsection{Context on Financial Markets and Algorithmic Traders}
\label{section:context}

Over the course of this project, there are a number of specialised terms and concepts relevant to our work, especially regarding the LOB, which are going to be expanded on in the following sub-section.

At the core of most financial markets lies the Continuous Double Auction (CDA) mechanism \cite{Smith1962Experimental}. Unlike the traditional auction setup, where items are sold one at a time with bidders actively competing until the highest price is reached, the CDA operates continuously, allowing buyers and sellers to place orders at any time. The 'double' in CDA signifies that it facilitates both buying and selling, dynamically matching buy orders with corresponding sell orders based on price preferences.

Central to the operation of the CDA is the LOB. The LOB is a dynamic, electronic record of all the outstanding buy and sell orders in the market for a particular asset. These orders are organised by price level, with the 'bid' price representing the maximum amount a buyer is willing to pay and the 'ask' price indicating the minimum amount a seller is willing to accept. The difference between the highest bid and the lowest ask is known as the 'spread.' A key feature of the LOB is that orders are processed based on price-time priority. This means that orders at the best price are always executed first, and among orders at the same price, the one placed earlier gets priority.
In a typical market scenario, traders—either humans or algorithms—submit orders. These orders can be of two main types:

\begin{itemize}
    \item Limit Orders: A trader specifies a price and quantity. For buyers, this price is the maximum they're willing to pay, and for sellers, it's the minimum they're willing to accept. These orders are added to the LOB and wait for a matching order to arrive.
    \item Market Orders: A trader specifies only the quantity, aiming to buy or sell immediately at the best available price. These orders are not added to the LOB; instead, they are matched with the best available opposite order from the LOB.
\end{itemize}

The market's primary objective is to facilitate trading by matching buy and sell orders. The continuous updating and matching in the LOB ensure liquidity and dynamic price discovery, reflecting the current consensus value of an asset.

The data we require from the LOB is referred to as ”Level-2” data, meaning that we get all the current active orders. For context, ”Level-1” market data contains only the prices and quantities of the best bid and ask in the market.

The Threaded Bristol Stock Exchange (TBSE) is an advanced, asynchronous version of the open-source Bristol Stock Exchange (BSE) \cite{2022GitHub}, a faithful, detailed simulation of a financial exchange where a variety of public-domain automated trading algorithms interact via a CDA. It is asynchronous in the way traders interact with the market, with each competing for an asset and placing orders concomitantly. Abiding by Smith's guidelines \cite{Smith1962Experimental}, traders solely aim for profit, ensuring no trades occur at a loss. Unlike its predecessor, where traders were sequentially and randomly polled for orders, TBSE grants each trader its own thread. Throughout a market session, traders continuously receive market updates and decide on placing orders. This structure privileges faster algorithms, as orders are queued on a "first in, first out" (FIFO) basis, emulating real-world market dynamics more closely.

The following terms will be relevant when defining our model's features. The LOB midprice is the average of the highest bid and the lowest ask prices in the LOB. The microprice refines this midprice by factoring in the order imbalance and the depth of the order book. Imbalance represents the proportionate difference between buy and sell orders, highlighting directional pressure. Total quotes on the LOB refer to the aggregate of all buy and sell orders present. The estimate $P*$ of the competitive equilibrium price predicts where supply meets demand, ensuring market clearance. Lastly, Smith's \emph{"alpha"} $\alpha$ metric gauges how closely the market price approaches this equilibrium, serving as a measure of market efficiency.

Now having cleared the domain-specific context, we transition to showing how experimental economics evolved from Smith's inaugural work to AI algorithmic traders, understanding how our work builds on existing knowledge in Section \ref{section:background}. The rest of this paper, based on \cite{cismaru2023}, will detail how the model that DTX uses was trained and the experimental setup in Section \ref{section:methods}. The results showing how DTX outperforms existing traders are shown in Section \ref{section:results}. Section \ref{section:discussion} will further analyse these findings, with Section \ref{section:lim-futwork} providing a view on limitations and future work, concluding with Section \ref{section:conclussion}. \cite{chatgpt}

\section{\uppercase{Background}}
\label{section:background}
\subsubsection{Beginnings of Experimental Economics and Agent Based Modelling}

The groundwork for experimental economics was laid by Vernon Smith in 1962 by publishing ”An Experimental Study of Competitive Market Behaviour” in The Journal of Political Economy (JPE) \cite{Smith1962Experimental}. Smith has implemented a series of experiments based on the CDA system, where buyers and sellers are announcing bids and others in real-time, with the possibility of a trade being executed any time the prices match.

The experiments were performed with small groups of human traders. They were instructed to trade an arbitrary commodity on an open-pit trading floor with the intention of maximising profitability, namely the difference between the limit price and the trade price. Each trader was given a pre-defined limit price: for sellers, the minimum they are allowed to sell their units at, and for buyers, the maximum price they can pay for a unit of the traded asset, thus preventing loss-making trades. The simulations were carried out as ”trading days”, namely time intervals of 5 to 10 minutes. The quotes that were shouted by the traders resembled the LOBs of modern markets. Once a trader agreed on a trade with its counterparty, both would leave the market as they only had a single unit to trade.
The results showed rapid convergence to the theoretical equilibrium price, measured by Smith’s $\alpha$ metric. It measures how well and efficiently the market is converging to the equilibrium price. The experiments capture the asynchronous nature of financial markets, one of the issues that this work is aiming to explore. Vernon Smith received the Nobel Prize in 2002 for his pioneering work in experimental economics, with his experiment styles being the basis of most research carried out in this field and the methodology used in this paper.

Three decades later, in 1993, Gode and Sunder introduced the Zero Intelligence traders \cite{Gode1993Allocative}. Their focus is on studying how automated traders perform in markets dominated by human traders. They introduced two trading strategies: Zero Intelligence Unconstrained (ZIU) and Zero Intelligence Constrained (ZIC). ZIU is generating purely random quotes, while ZIC is limited, constrained to a price interval. Their experiments, carried out in the style of Vernon Smith, showed ZIC to outperform human traders. A few years later, in 1997, Cliff published a paper proposing Zero Intelligence Plus (ZIP) traders, which, by using a simple form of ML, can be adaptive and converge in any market condition \cite{cli1997minimal}. ZIP is based on a limit price and an adaptive profit margin. The margin is influenced by a learning rule and the conditions of the market.

In 1998, Gjerstad \& Dickhaut described an adaptive agent, GD \cite{Gjerstad1998Price}, with Tesauro \& Bredin publishing a paper in 2002 describing the GD eXtended (GDX) trading algorithm \cite{tesauro2002strategic}. In 2006, Vytelingum’s thesis introduced what is called the Aggressive-Adaptive (AA) strategy \cite{vytelingum2006structure}, which was thought to be the best-performing agent until recently. In 2019, Cliff and Snashall performed comprehensive experiments comparing AA and GDX, simulating over a million markets. The results show that AA is routinely outperformed by GDX, arguing that advancements in cloud computing and compute power open new possibilities for strategy evaluation that were not possible before \cite{Snashall2019Adaptive}.

\subsubsection{Rise of Intelligence in Market Modelling and Price Prediction}

The advent of AI has attracted increased attention in the fields of finance and trading. More and more papers detail how advanced Deep Learning methods became very powerful tools in the world of agent-based trading, market making, and price forecasting. Axtell and Farmer present in their 2018 report how advances in computing have enabled agent-based trading (ABM), which has impacted how trading is performed today \cite{axtell2022agent}. In finance, ABM helped us understand markets, volatility, and risk better. Their report is comprehensive and can be considered a higher-level point of reference on how agents are being applied in different branches of finance and economics. Njegovanovi{\'c} published a paper in 2018 that discusses the implications of AI and ML in finance, with a focus on how the human brain and its behaviour have inspired the architecture of automatic decision models \cite{njegovanovic2018artificial}.

In the past decade, a number of studies have explored the potential of Machine Learning (ML) and Deep Learning in finance. In 2013, Stotter, Cartlidge, and Cliff introduced a new method for assignment adaptation in ZIP, performing balanced group tests against the well-known ZIP and AA strategies \cite{stotter2013exploring}. Their results show that assignment-adaptive (ASAD) traders equilibrate more quickly after market shocks than base strategies. In 2019, Ji, Kim, and Im performed a comparative study of DNN vs. LSTM for Bitcoin price prediction, trained on historical data from public ledger records. They conclude that classification models (DNN) perform better than regression models (LSTM) for price prediction \cite{ji2019comparative}.

Another paper from 2020 by Silva, Li, and Pamplona uses LSTM-based trading agents to predict future trends in stock index prices. Their proposed method, named LSTM-RMODV, demonstrates the best performance out of all studied methods, and it is shown to work in both bear and bull markets \cite{silva2020automated}. The results found Deep Reinforcement Learning to be performant in market-making applications by Sun, Huang, and Yu in 2022 \cite{Sun2022Market}, with a similar piece of work being published in 2019 \cite{Sirignano2019Universal}. They propose a Deep Learning model applied to historic US equity markets. The information extracted from the LOBs uncovers a relationship between past orders and the direction of future prices. They conclude that this is better than specialised predictions for specific assets. Their results illustrate the applicability and power of Deep Learning methods in modelling market behaviour and generalisation.

\subsubsection{Need for Intelligence and Realistic Modelling}

The work in this paper continues what Calvez and Cliff started in 2018 \cite{Calvez2018Deep}, when they introduced a DLNN system trained to replicate adaptive traders in a simulated market. Purely based on the observation of the best bid and ask prices, the DLNN has managed to perform better than the trader observed. In 2020, Wray, Meades, and Cliff will take this further by introducing the first version of DeepTrader, a high-performing algorithmic trader \cite{Wray2020Automated} trained to perform in a sequential market. Based on a LSTM, it automatically replicates a successful trader by training on 14 features derived from Level-2 market data. The first version of DeepTrader matches or outperforms existing trading algorithms in the public-domain literature.
Most studies are performed on sequential simulations, in which the speed at which the traders react to changes in the market does not matter. Axtell and Farmer argue in their report ”Agent-Based Modelling in Economics and Finance: Past, Present, and Future”, mentioned above \cite{axtell2022agent}, that the real social and economic worlds are parallel and asynchronous, but we try to replicate it with single-threaded code. Rollins and Cliff try to mitigate this in a paper they published in 2019 \cite{rollins2020trading}. They propose TBSE, as we introduced in Section \ref{section:context}, on which they perform pair-wise experiments between well-known trading strategies. The results reported intriguing insights, with a new dominance hierarchy of trading algorithms emerging, opening a new area of research. 

Our work aims to build a new strategy to trade in the TBSE, leveraging the advantages of the architecture of DeepTrader. We will dive into the details of this in the next section, asking the question: Can we extend this model to learn from a variety of traders and study its behaviour in a parallel simulation? We hope that the results of our study can provide more insight into its potential real-world performance.

\section{\uppercase{Methods}}
\label{section:methods}

The core of our work relies on TBSE, as introduced in Section \ref{section:context}. It was used to generate the large amounts of data required for training the LSTM network used by DTX when running against the legacy trading strategies. The code of our project is available online at \href{https://github.com/armandcismaru/DeepTraderX}{GitHub at github.com/armandcismaru/DeepTraderX} for easy reproducibility.

TBSE was designed to use real-world historical data to introduce variability in its supply and demand schedules, generating more realistic data. The simulator uses a stochastically-altered schedule at each session, avoiding repetition and bias. For our training and experiment sessions, we used IBM stock price data from the August 31, 2017 NYSE trading day.

The simulator provides the means to produce large quantities of "historical" market data. The one metric we are interested in is the profit that each trader type achieves at the end of the session, namely profit per trader (PPT). We cannot assess these algorithmic traders the same way we would with real traders, as TBSE doesn't simulate loss, so we are judging based on profits only.

The data used as input to the model is curated by taking snapshots of the Level-2 LOB data, updated each time a trade occurs. We use 14 multivariate input features as training data for our DLNN-based model, deriving from these LOB snapshots, as detailed in Section \ref{section:context}. When performing inference, our model uses this data to produce the target variable, namely the price at which it is willing to trade at a specific time in the market (the quote placed by the trader). The 14 features are as follows:

\begin{enumerate}
    \item The time $t$ of the trade when it took place.
    \item The type of customer order used to initiate the trade, either a "bid" or an "ask" order.
    \item The limit price of the trader's quote that initiated the trade.
    \item The midprice of the LOB at time $t$.
    \item The microprice of the LOB at time $t$.
    \item The LOB imbalance at time $t$.
    \item The spread of the LOB at time $t$.
    \item The best (highest) bid on the LOB at
    time $t$.
    \item The best (lowest) ask on the LOB at time $t$.
    \item The difference between the current time and the time of the previous trade.
    \item The quantity of all quotes on the LOB at time $t$.
    \item An estimate $P^*$ of the competitive equilibrium price.
    \item Smith’s $\alpha$ metric using the $P^*$ estimate of the competitive equilibrium price at time $t$.
    \item \emph{The target variable}: the price of the trade.
\end{enumerate}

As in real markets, when trading live in the simulation, DTX will only have access to its own limit price, w.r.t. feature number 3. The data used for training includes each trader's own limit price.

\subsubsection{Data Generation and Preprocessing}

TBSE provided five working trading agents that were used to generate the training data, as included here: \href{https://github.com/MichaelRol/Threaded-Bristol-Stock-Exchange}{github.com/MichaelRol/Threaded-Bristol-Stock-Exchange}. In order to diversify the data and cover a lot of market scenarios, the simulations were run using different proportions of the trading strategies, adding to a total of 40 traders per simulation. The following proportions of 20 traders per side of the exchange (buyers or sellers) were used to generate the training data: (5, 5, 5, 5, 0), (8, 4, 4, 4, 0), (8, 8, 2, 2, 0), (10, 4, 4, 2, 0), (12, 4, 2, 2, 0), (14, 2, 2, 2, 0), (16, 2, 2, 0, 0), (16, 4, 0, 0, 0), (18, 2, 0, 0, 0), and (20, 0, 0, 0, 0). Each number in a specific position corresponds to a population of traders of a certain type for a market simulation. For example, for the specification (12, 4, 2, 2, 0), there are 12 ZIC, 4 ZIP, 2 GDX, 2 AA, and no Giveaway traders for both the buyers and sellers sides.

Using all the unique permutations of the schedules resulted in 270 trader schedules, ensuring that the traders were used evenly. Each schedule was executed for 44 individual trials, amounting to $270 \times 44 = 11880$ market sessions. Each session represented an hour of simulated market time, requiring a bit over one minute of real wall-clock time. Running on a single computer, generating this amount of data would require approximately 8.6 days of continuous execution. Given this time constraint, the decision was made to use cloud computing.

The data generation system was made possible by running the code as Docker containers, with the computation parallel distributed across 10 AWS \emph{Elastic Compute Cloud} (EC2) instances. The workload was split equally amongst the virtual machines (VMs) by \emph{Kubernetes}, with their provisioning done by AWS's \emph{Elastic Kubernetes Service} (EKS).

The files generated amount to roughly 13 million LOB snapshots, one per line. In order to save time and prepare the data for training, the Python \emph{Pickle} library was used to serialise the CSV files to a large byte stream file.

It is generally good practice to normalise the inputs of a network due to performance concerns, particularly for Deep Learning architectures like LSTMs. Normalising the inputs helps ensure that all features are contained within a similar range and prevents one feature from dominating the others. For example, we have features with different scales, such as the time, which runs from 0 to 3600, while the quote type is binary. So by normalising, we only have values in the [0,1] interval. Doing this ensures improved convergence of the optimisation algorithm and helps the model generalise better to new data. The choice was to use min-max normalisation, given that we are working with multivariate features derived from financial data, so it is important to preserve their scale while using an easy-to-understand model.

\begin{figure*}[!ht]
  \centering
   {\epsfig{file = 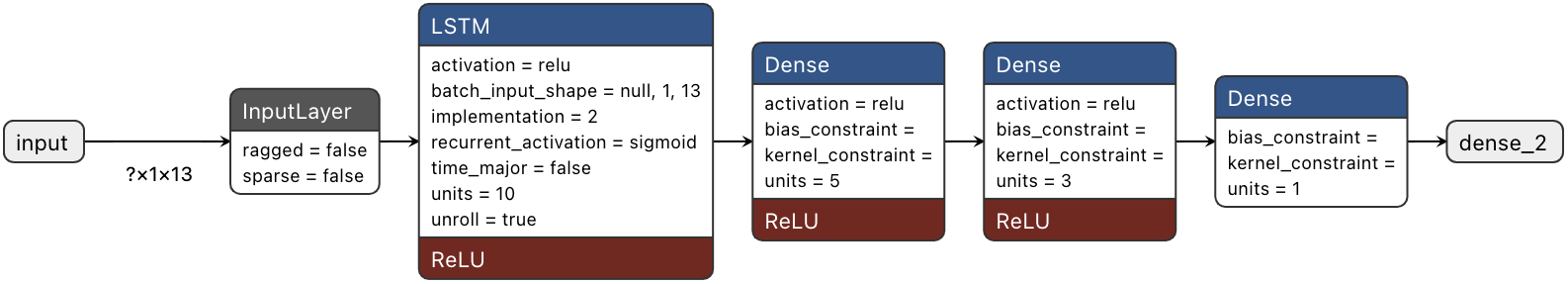, width = 0.9\linewidth}}
  \caption{Architecture diagram for the DLNN model used by DTX.}
  \label{fig:architecture}
\end{figure*}

\subsubsection{Model Architecture and Training}
Contrary to the usual practices for training and validating a DLNN, which consist of splitting the dataset into \emph{training}, \emph{validation}, and \emph{test} subsets, we used all the dataset for training. Markets are a combination of unique factors, so our trader's profit is heavily dependent on what is happening in a specific simulation. Considering this, it is without purpose to assess its performance relative to historic data by judging the absolute values of our target variable. Rather, as the model produced a good drop in the loss level during training, the DLNN was validated by quantifying how well DTX performed in live market simulations against other traders in terms of PPT. Our dataset is large and was generated using unique simulations; thus, DTX doesn't learn to replicate specific scenarios; rather, it grows its ability to adapt and generalise in any condition.

The model DTX relies on is illustrated in the architecture diagram in Figure \ref{fig:architecture}. It is comprised of three hidden layers, a LSTM with 10 units (neurons), and two consequent Dense layers with 5 and 3 units, respectively, all using the Rectified Linear Unit (ReLU) activation function. The output layers use a ”linear” activation function, chosen as suitable for a continuous output variable.

When dealing with large datasets, training should be done in batches in order to accommodate memory limitations and speed up training. Our network accommodates this by having a custom data generator that implements the Sequence class, which Keras uses to train a model in batches. To balance accuracy and training times, a batch size of 16384 was chosen. The learning rate of $\mu=1.5\times10^{-5}$ was deemed the sensible choice, trying to balance potential overfitting and long convergence times. The DLNN uses the Adam optimizer for its ability to efficiently converge to a good solution, prevention of overfitting, and incorporation of momentum, which helps speed up learning and improve generalisation performance.

To speed up the training, the LSTM layer uses unrolling, which processes input sequences in parallel instead of sequentially. During training, 28 worker nodes were used to pre-load the batches in memory to reduce computational overhead. The network was trained using the \emph{Blue Crystal 4} supercomputer, leveraging the power of its powerful GPUs, which Tensorflow is designed to optimise on. The training time required is approximately 22 hours.

The model was trained for 20 epochs. An epoch refers to a single pass of the entire dataset through the neural network. During each epoch, the model is exposed to each point of the dataset once. So, at the end of the training session, DTX gets exposure to $11880 \times 20=237600$ market sessions. With each experiment producing $\sim1100$ LOB snapshots, DTX is trained using a total of roughly 261 million snapshots. The training error (loss) was calculated using the Mean Square Error (MSE). The error decreased considerably during the first 4 epochs and approached 0 in the last epoch with a loss curve rather resembling an asymptote to the X axis. 

\subsubsection{Experiment Design and Evaluation}

Finding the right methodology for comparing trading strategies is as important as the strategies themselves, as it is essential to isolate market conditions in repeatable experiments. Traders are dependent on the behaviour of other strategies, so this mandates the need to study them in a controlled environment, allowing quantitative analysis of their performance. Drawing inspiration from the work of Tesauro and Das \cite{tesauro2001high}, we have chosen two testing methodologies. 

The first experiment design is the Balanced Group Tests (BGTs), in which the buyer and seller populations are evenly split between two types of strategies. The choice of balanced-group tests provides benefits to our research, namely that it is a stochastic-controlled trial method that helps reduce bias sources and improve the internal validity of the study. We want to make sure that the differences observed come from differences between trading strategies, not noise. The tests allow full control of the experiment's conditions. The time frame of the simulation, the supply and demand schedules, and the order interval can all be controlled to isolate the differences between the chosen strategies. The second type of experiment is the One to Many tests (OTMs), where the trading strategy that you want to observe becomes the "defector" out of a homogenous population made up of different strategies. This is useful for testing how an algorithm behaves when faced with defection and invasion. For fairness, there is one defecting strategy on both buyer and seller sides.

The research on the profitability of DTX has been conducted against four "competitor" traders: ZIC, ZIP, GDX, and AA, adding up to eight sets of head-to-head experiments. These strategies were chosen as they are the most relevant in the literature, with AA, GDX, and ZIP being "super-human" traders, amongst the first to be proven to outperform humans. 

For each experiment, the trained model that DTX uses has been run in $n=500$ independent market simulations. It is worth specifying that each set of 50 trials was run on a different cloud machine, resulting in a broad distribution of profits, with each set of 50 experiments using the same seed for functions involving randomness. This is due to the well-known and researched issue in computer science that machines cannot emulate perfect randomness \cite{Bridle2022Crucial}.

\section{\uppercase{Results}}
\label{section:results}

The following section presents the results of our experiments, obtained through 4,000 individual market simulations. The outcome largely supports our research hypothesis, with DTX dominating in 6 out of 8 experiments, with very significant differences in PPT for a number of them. 

Due to space constraints, we have limited the choice of graphic support to profit distribution box plots and scatter plots of individual trials. An extended, different summary of the results can be found in Chapter 3.8 of \cite{cismaru2023}. In the box plots, the vertical axis is represented by PPT across trials. The box represents the interquartile range, the range between the first quartile and the third quartile. The line inside the box represents the median of the dataset. The whiskers represent the data within 1.5 times the interquartile range, with the diamond-shaped points outside them being considered outliers from a data distribution point of view. The scatter plots show individual trials in terms of PPT obtained by both traders. The line in the scatter plot is a diagonal reference line, where the points would lie if the profits per trader for both strategies were equal. Points above the line indicate higher profits achieved by DTX and are analogous for the other trader.

The figures are grouped on an opponent basis, with each set of two box-plots showing the PPT distributions of the BGTs and OTMs, followed by their corresponding scatter plots. We present them in the following order: ZIC, ZIP, GDX, and AA. Chapter 4 in \cite{cismaru2023} also provides an extensive description of the statistical significance tests conducted for each experiment. Thus, for each time DTX was poised against the other traders, we performed a Wilcoxon-signed rank-test with a significance level of $95\%$. The null hypothesis is that there is no statistical difference between the means of the profits achieved by the traders. A \emph{p-value} lower than $0.05$ indicates that we can reject the hypothesis, concluding that one strategy outperforms the other in a given experiment.

\subsubsection{ZIC vs. DTX}

\begin{figure}[htbp]
    \centering
    \begin{subfigure}{0.494\columnwidth}
        \centering
        \includegraphics[width=1\linewidth]{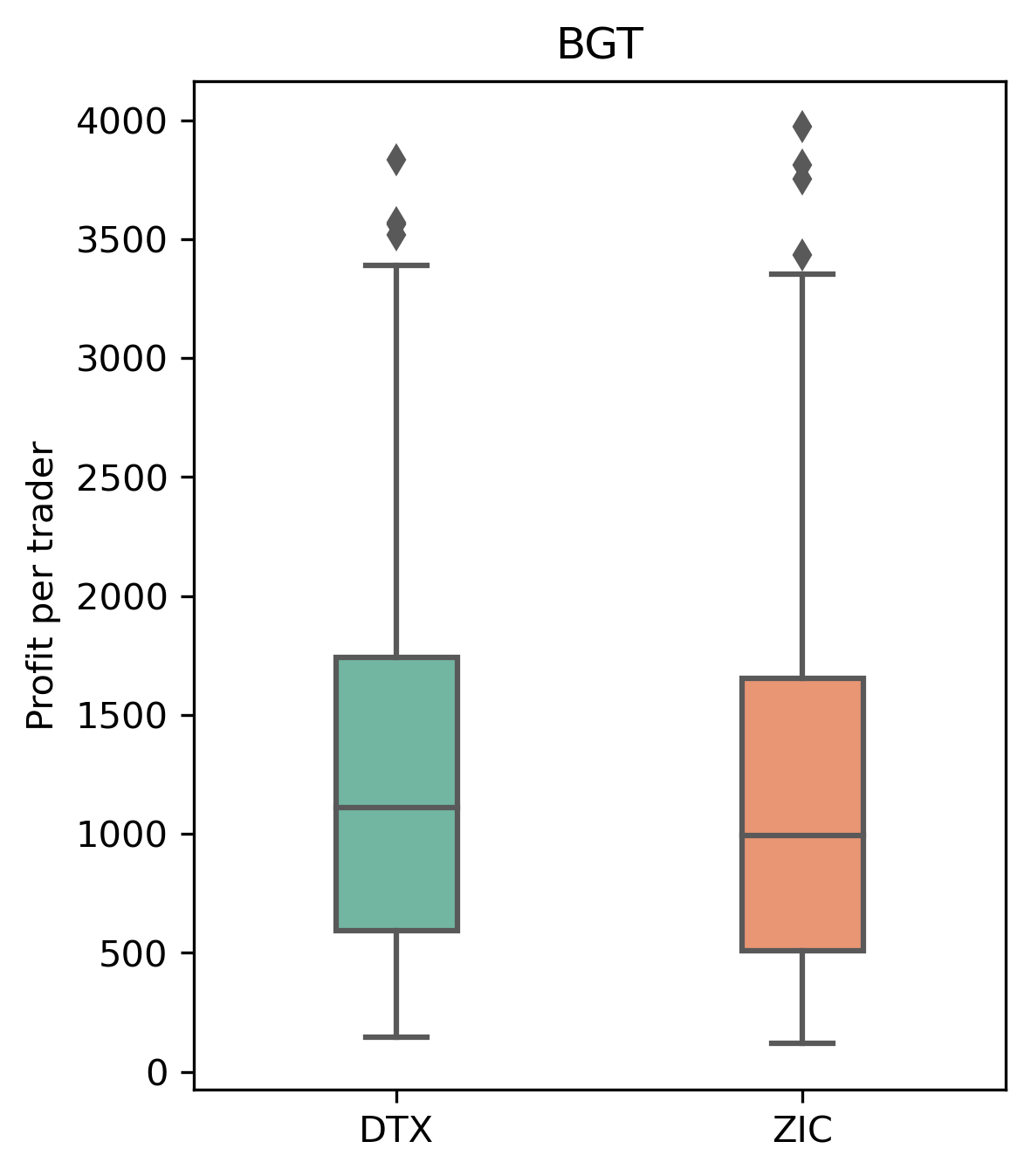}
        \caption{Balanced Group Tests for ZIC vs. DTX.}
        \label{fig:zic-bgr-dist}
    \end{subfigure}
    \begin{subfigure}{0.494\columnwidth}
        \centering
        \includegraphics[width=1\linewidth]{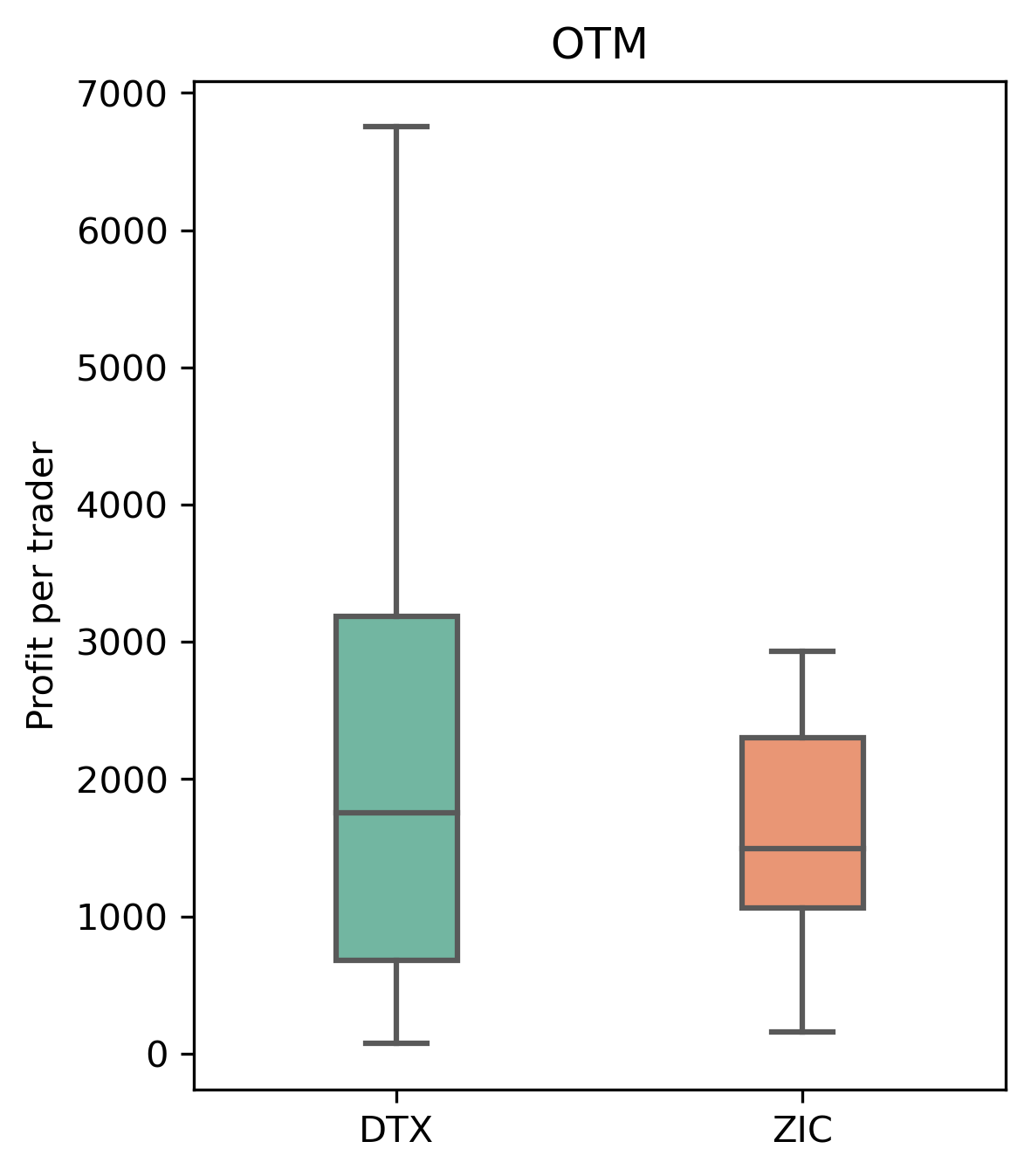}
        \caption{One to Many Tests for ZIC vs. DTX.}
        \label{fig:zic-otm-dist}
    \end{subfigure}
    \vspace{1pt}
    \caption{Box-plots showing PPT for ZIC vs. DTX tests.}
    \label{zic-dist:image}
\end{figure}

\begin{figure}[!ht]
  \centering
   {\epsfig{file = 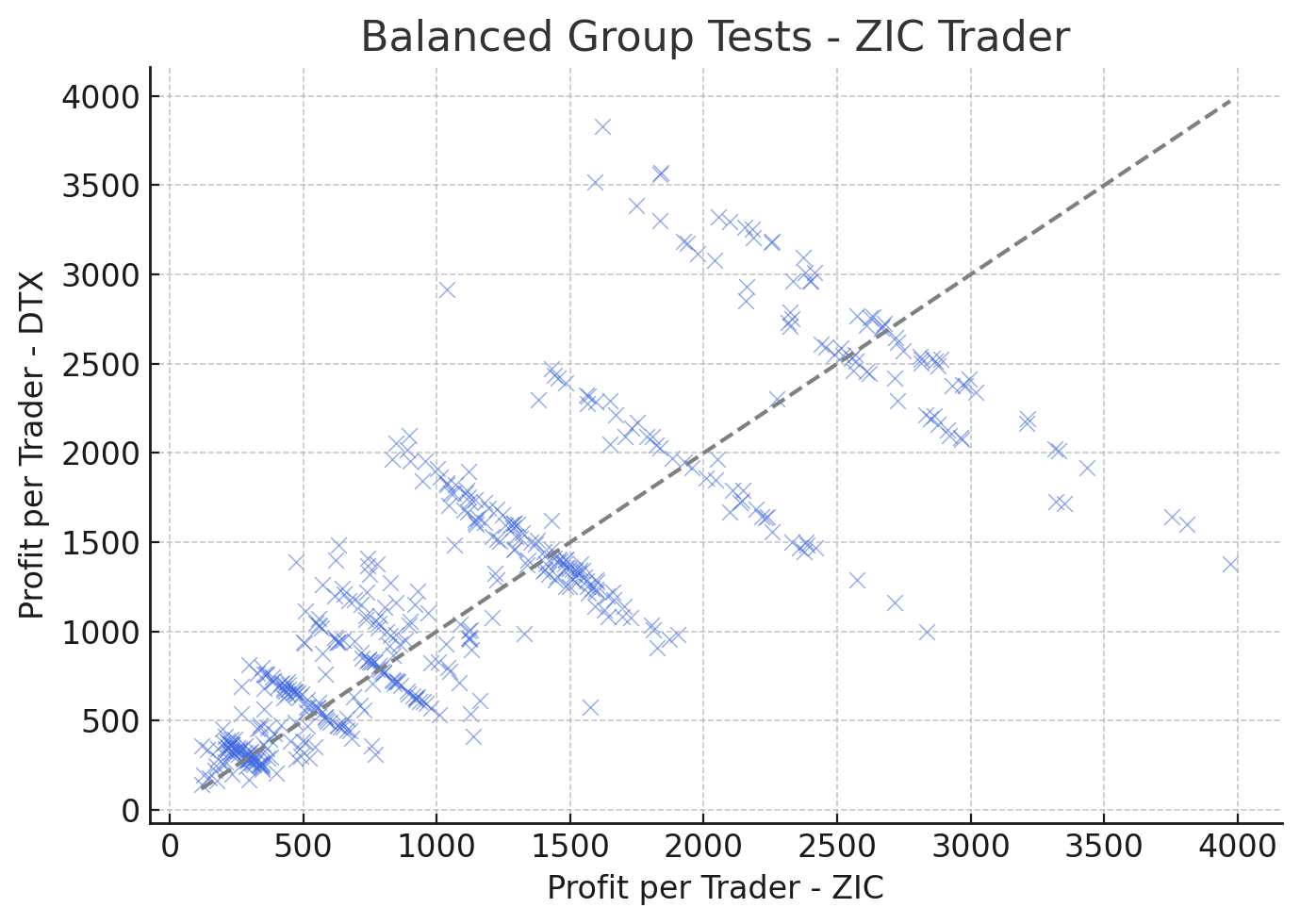, width = 7.5cm}}
  \caption{Scatter plot of PPT in BGTs for ZIC vs. DTX.}
  \label{fig:bgr-zic-dtx}
\end{figure}

\begin{figure}[!ht]
  \centering
   {\epsfig{file = 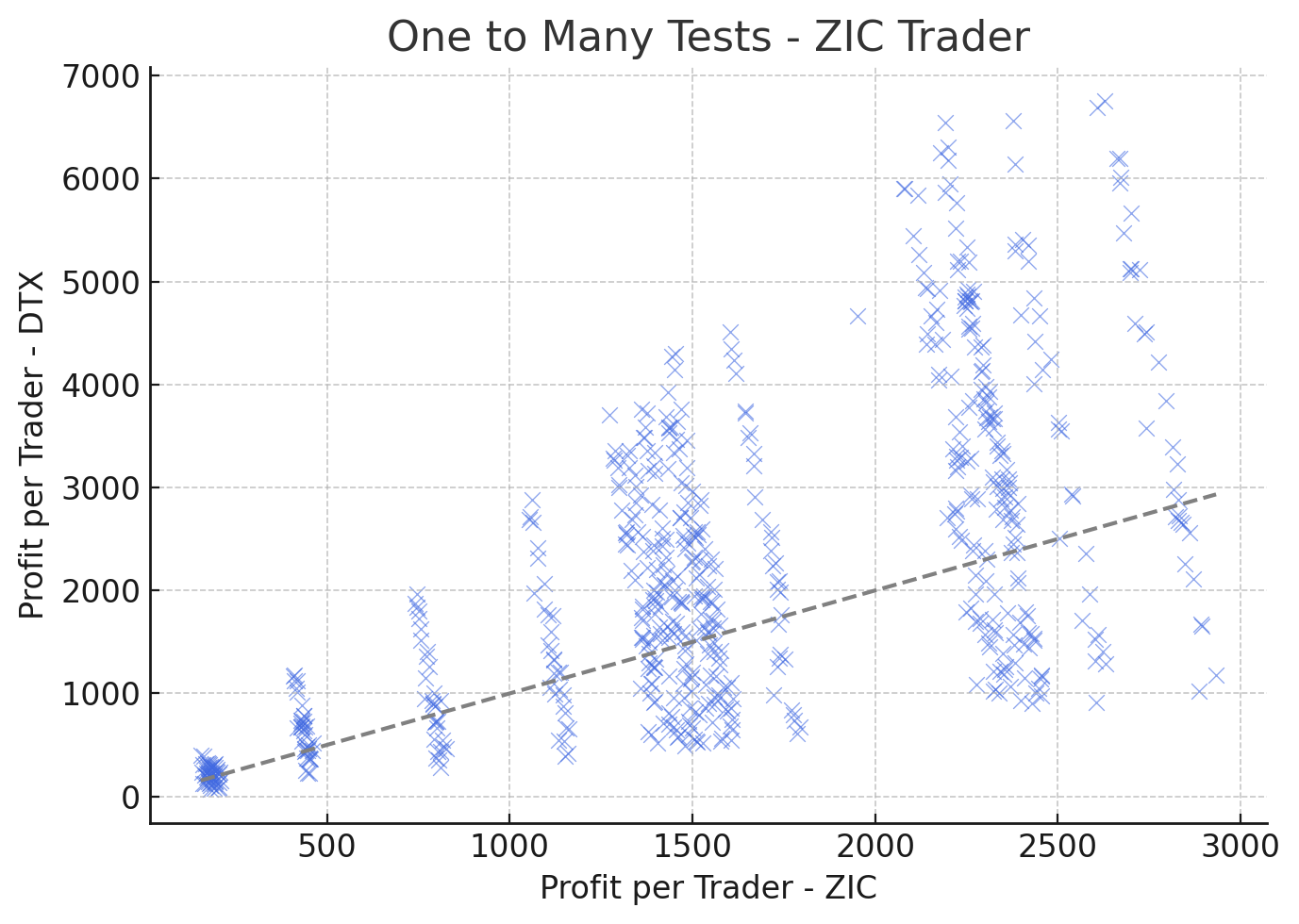, width = 7.5cm}}
  \caption{Scatter plot of PPT in OTMs for ZIC vs. DTX.}
  \label{fig:otm-zic-dtx}
\end{figure}

Figure \ref{fig:zic-bgr-dist} shows a narrow difference in means between ZIC and DTX in the BGTs, a fact visible by the dispersion of profits in Figure \ref{fig:bgr-zic-dtx}. The statistical test for 95\% significance level has confirmed DTX as the dominant strategy of this experiment. In the case of OTMs, the difference in profits is more sensible in favour of DTX, supported by the profit distribution in Figure \ref{fig:zic-otm-dist} and by the cluster of profits above the diagonal in Figure \ref{fig:otm-zic-dtx}. The statistical test has confirmed the significant dominance of DTX in this experiment.

\subsubsection{ZIP vs. DTX}

\begin{figure}[htbp]
    \centering
    \begin{subfigure}{0.494\columnwidth}
        \centering
        \includegraphics[width=1\linewidth]{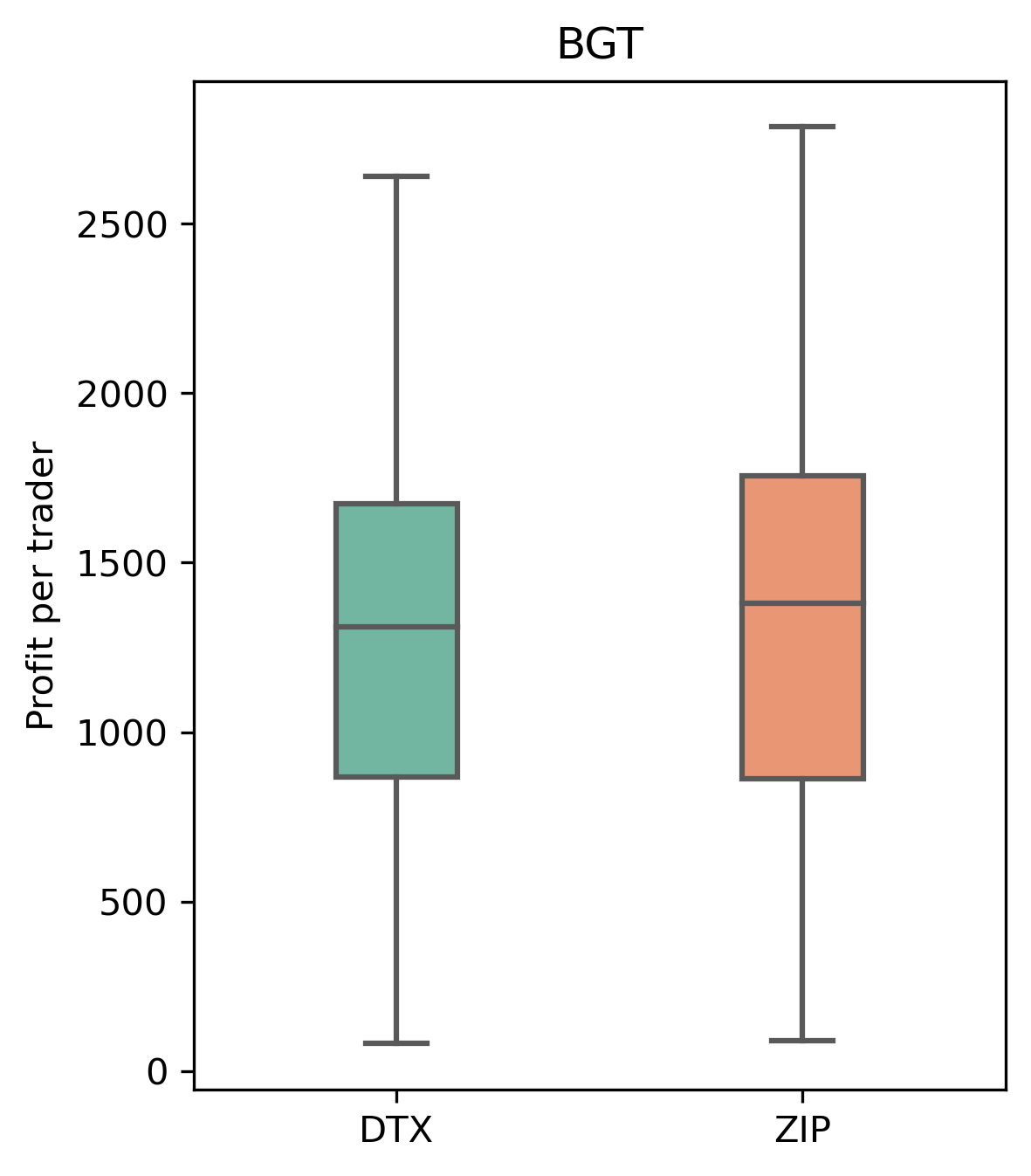}
        \caption{Balanced Group Tests for ZIP vs. DTX.}
        \label{fig:zip-bgr-dist}
    \end{subfigure}
    \begin{subfigure}{0.494\columnwidth}
        \centering
        \includegraphics[width=1\linewidth]{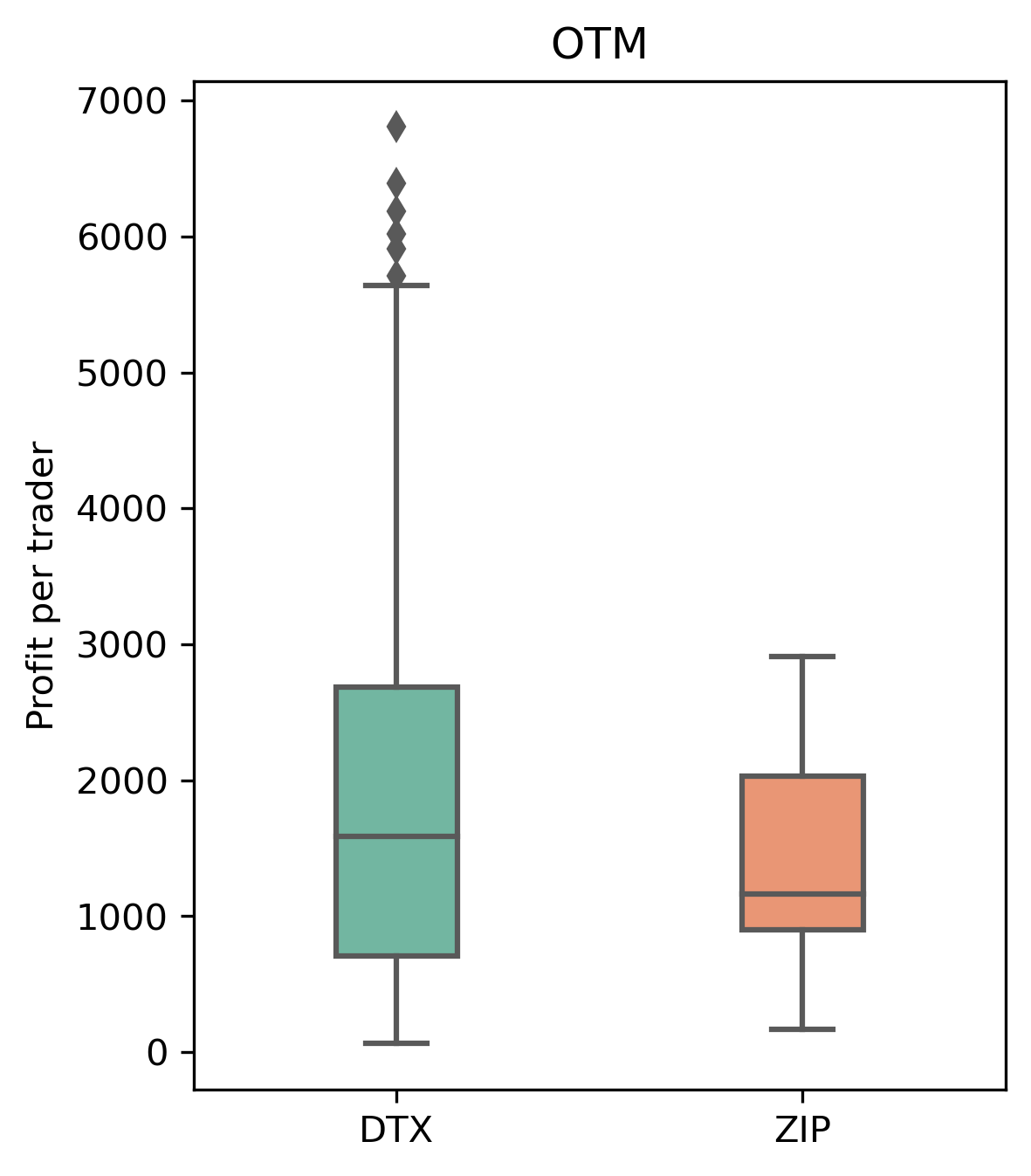}
        \caption{One to Many Tests for ZIP vs. DTX.}
        \label{fig:zip-otm-dist}
    \end{subfigure}
    \vspace{0.1cm}
    \caption{Box-plots showing PPT for ZIP vs. DTX tests.}
    \label{zip-dist:image}
\end{figure}

\begin{figure}[!ht]
  \centering
   {\epsfig{file = 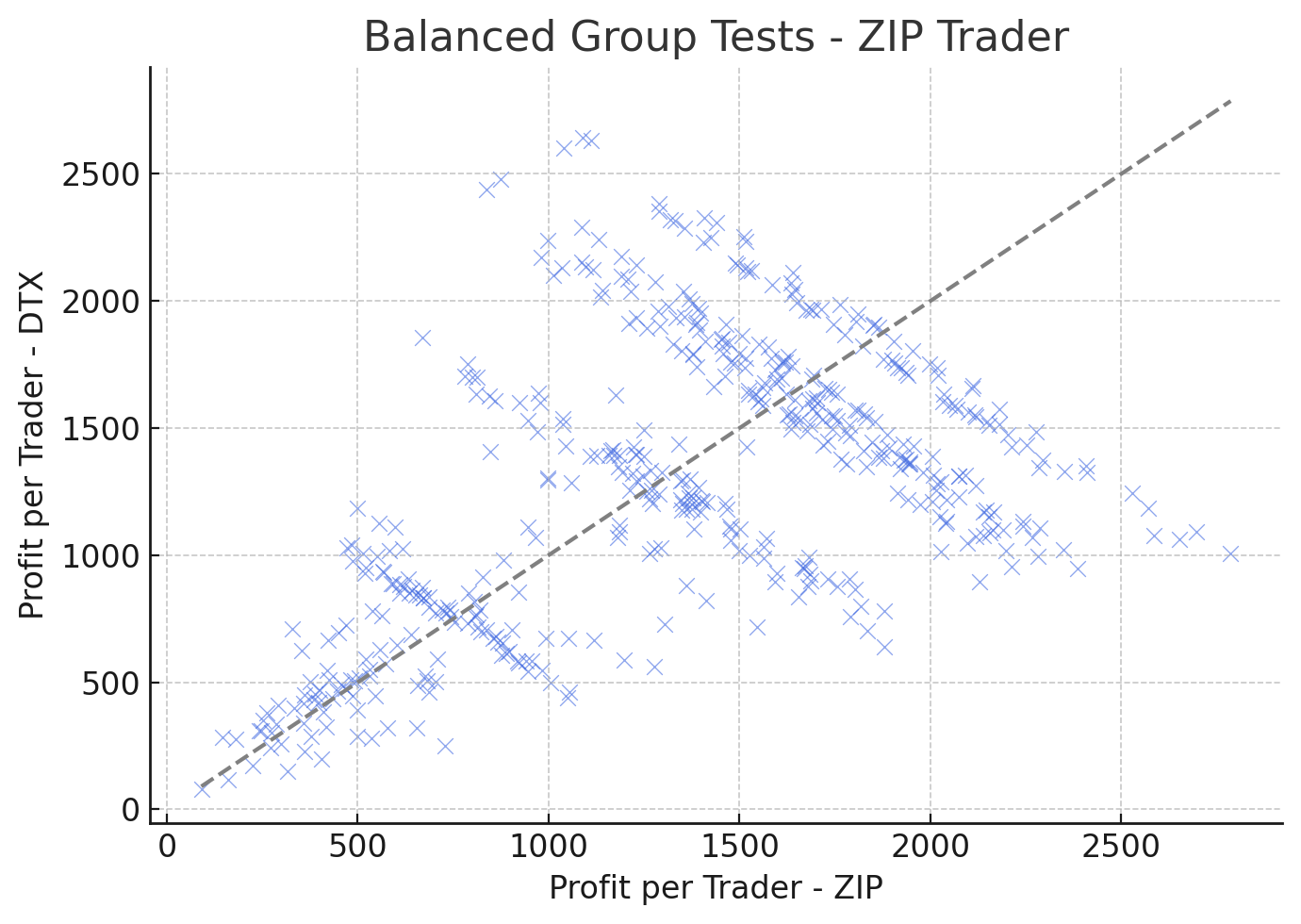, width = 7.5cm}}
  \caption{Scatter plot of PPT in BGTs for ZIP vs. DTX.}
  \label{fig:bgr-zip-dtx}
\end{figure}

The BGT experiment between ZIP and DTX is our model's only categorical loss. While Figure \ref{fig:bgr-zip-dtx} doesn't indicate any immediate winner, Figure \ref{fig:zip-bgr-dist} shows a slight advantage for ZIP, with a higher mean PPT. The Wilcoxon signed-rank test has confirmed the result, confirming that there is a significant difference in profits in favour of ZIP.

\begin{figure}[!ht]
  \centering
   {\epsfig{file = 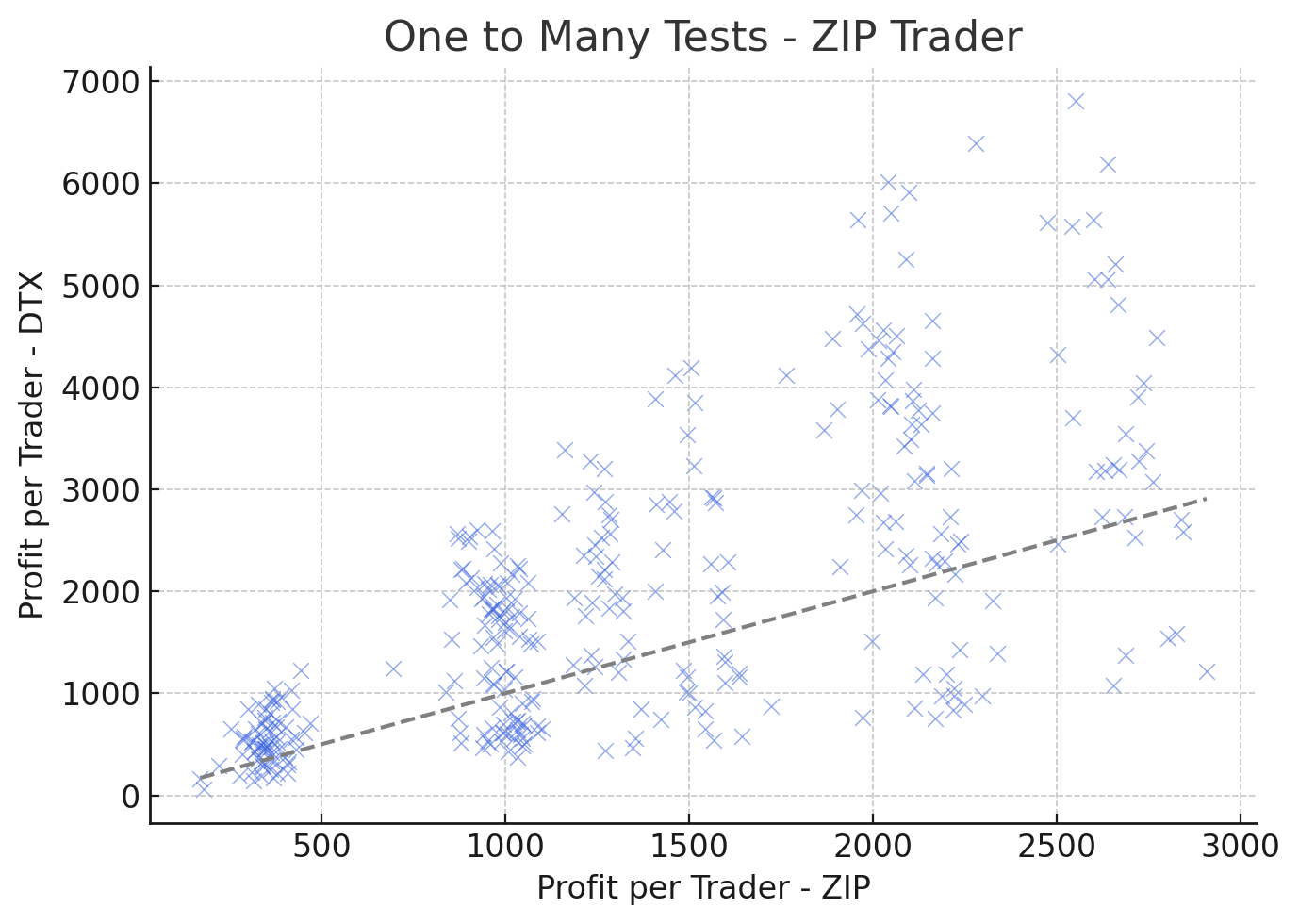, width = 7.5cm}}
  \caption{Scatter plot of PPT in OTMs for ZIP vs. DTX.}
  \label{fig:otm-zip-dtx}
\end{figure}

On the other hand, Figure \ref{fig:zip-otm-dist} shows higher mean profits for DTX, although with a much bigger variance and a number of outlier values. Visual inspection of the points in Figure \ref{fig:otm-zip-dtx} suggests that DTX significantly outperforms, a fact backed by the result of the statistical test.

\subsubsection{GDX vs. DTX}

\begin{figure}[htbp]
    \centering
    \begin{subfigure}{0.494\columnwidth}
        \centering
        \includegraphics[width=1\linewidth]{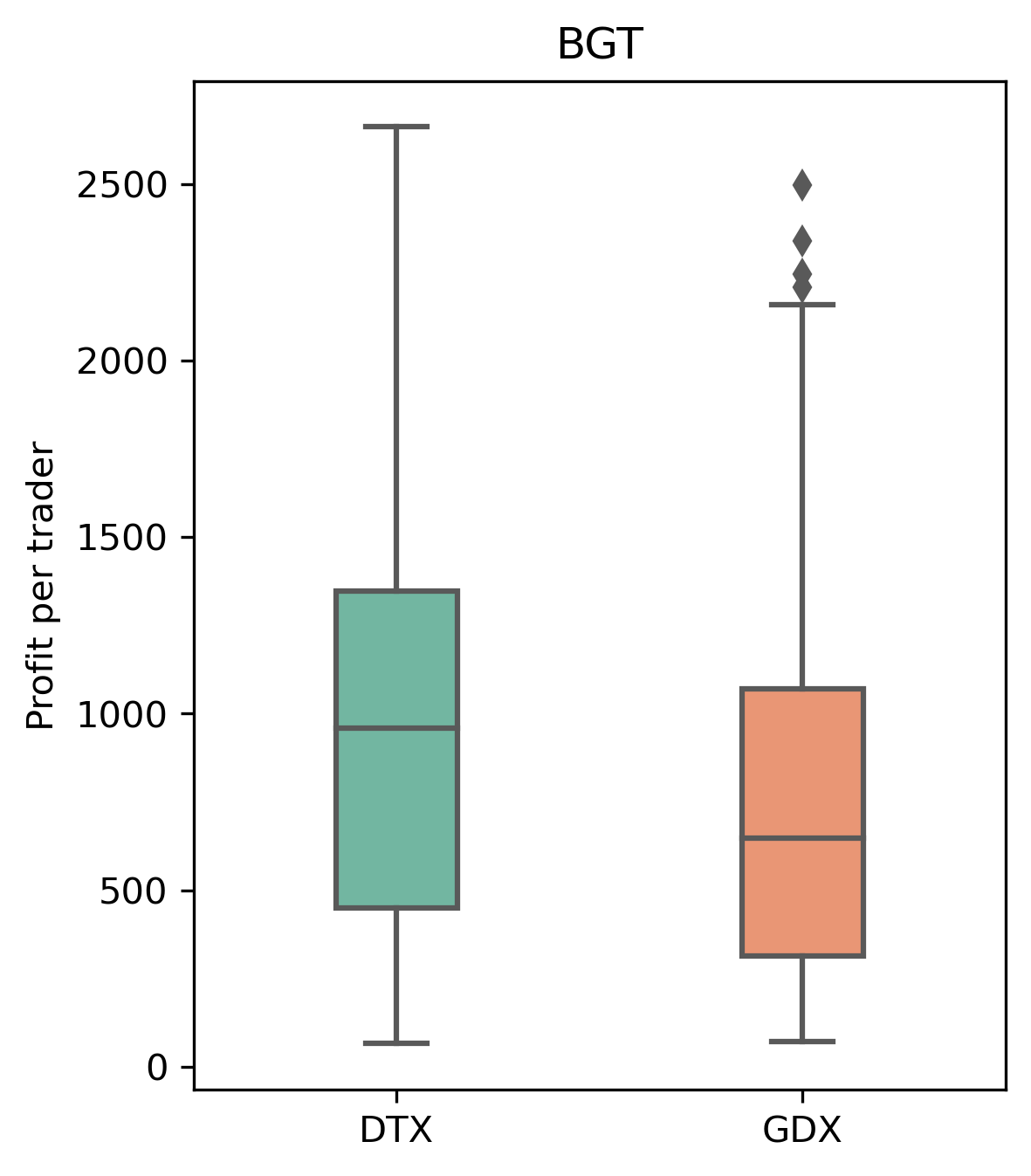}
        \caption{Balanced Group Tests for GDX vs DTX.}
        \label{fig:gdx-bgr-dist}
    \end{subfigure}
    \begin{subfigure}{0.494\columnwidth}
        \centering
        \includegraphics[width=1\linewidth]{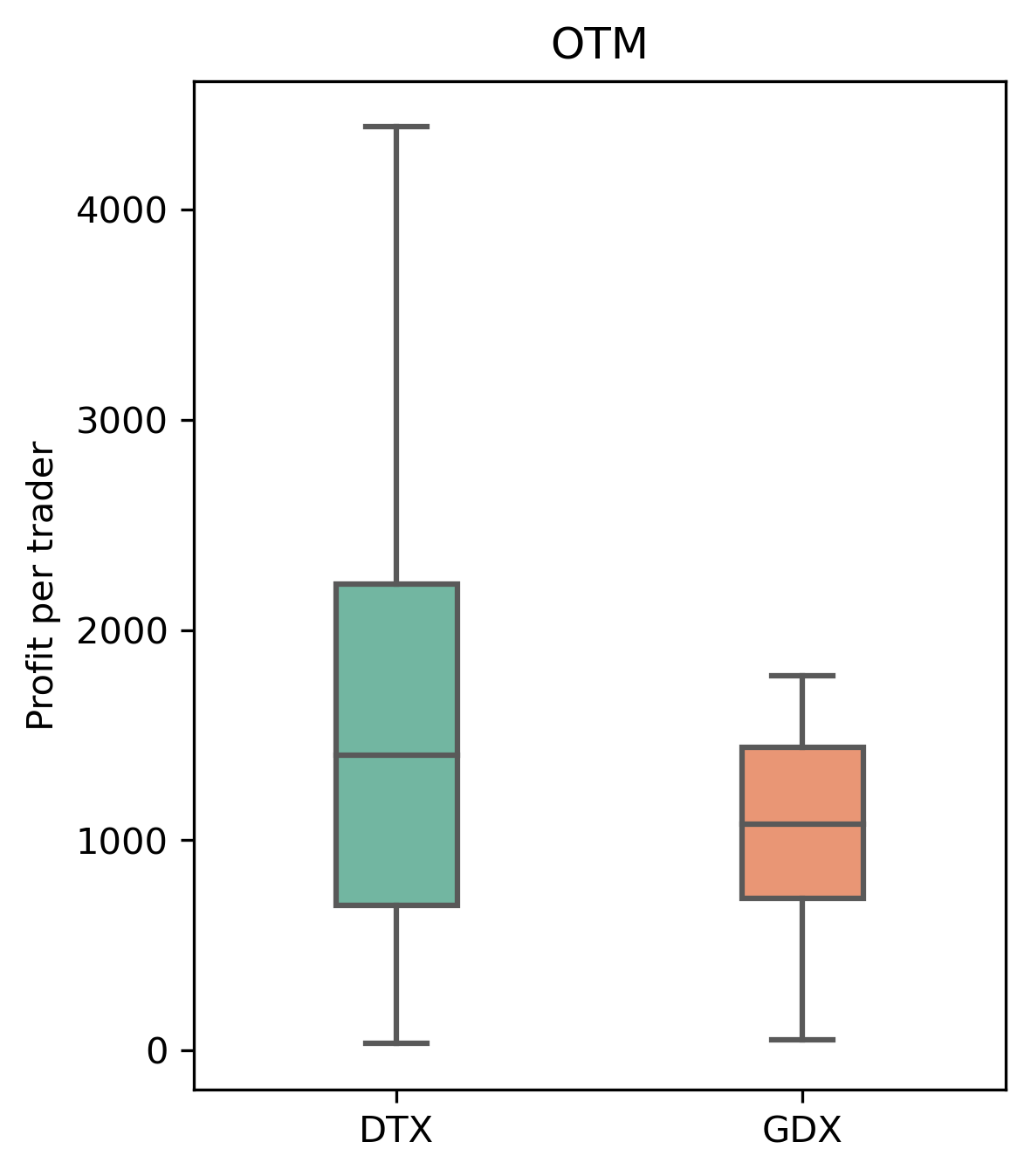}
        \caption{One to Many Tests for GDX vs. DTX.}
        \label{fig:gdx-otm-dist}
    \end{subfigure}
    \vspace{0.1cm}
    \caption{Box-plots showing PPT for GDX vs. DTX tests.}
    \label{gdx-dist:image}
\end{figure}

\begin{figure}[!ht]
  \centering
   {\epsfig{file = 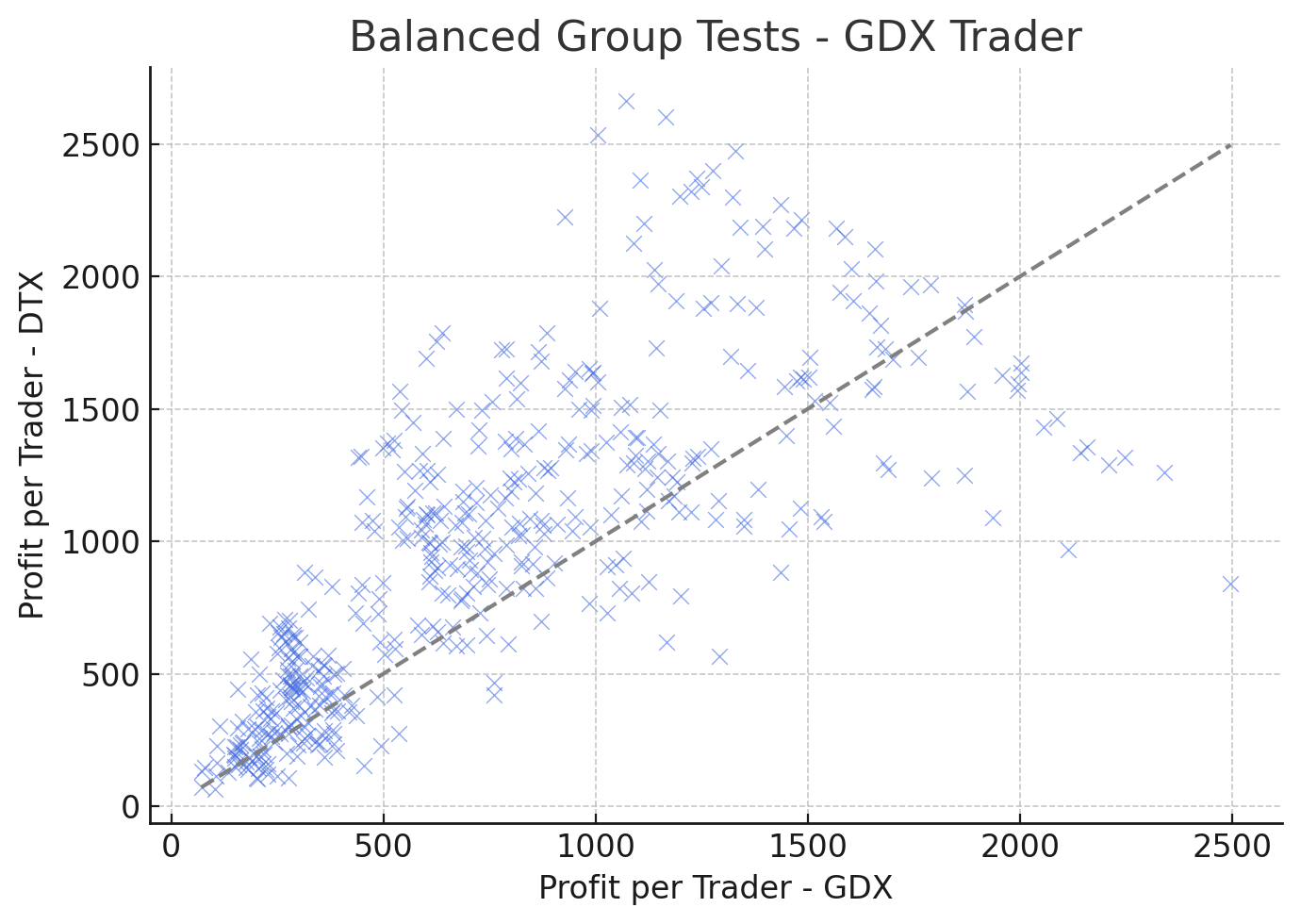, width = 7.5cm}}
  \caption{Scatter plot of PPT in BGTs for GDX vs. DTX.}
  \label{fig:bgr-gdx-dtx}
 \end{figure}

Figures \ref{fig:gdx-bgr-dist} and \ref{fig:bgr-gdx-dtx} show BGT comparison of PPT scores between DTX and GDX. Upon visual inspection, the bar plot shows a significant difference between the means of DTX and its competitor, with the scatter plot placing most of its points above the diagonal, indicating the clear dominance of DTX in this experiment, with the same outcome confirmed by the outcome of our statistical test.

\begin{figure}[!ht]
  \centering
   {\epsfig{file = 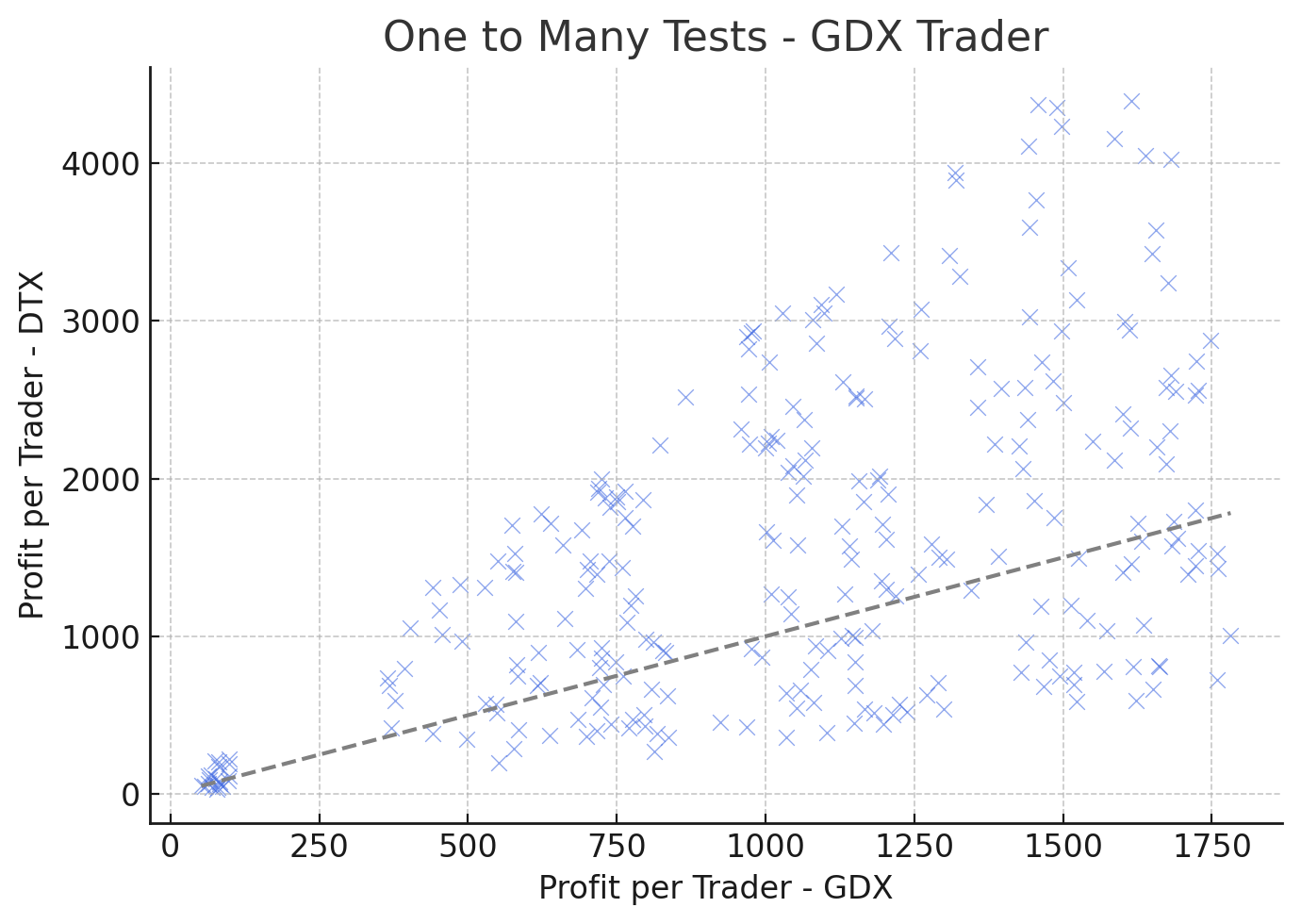, width = 7.5cm}}
  \caption{Scatter plot of PPT in OTMs for GDX vs. DTX.}
  \label{fig:otm-gdx-dtx}
\end{figure}

Figures \ref{fig:gdx-otm-dist} and \ref{fig:otm-gdx-dtx} show PPT score comparisons in the OTM experiment. The profits obtained by DTX, although they are dispersed and have high variance, lay on a superior magnitude scale than those of GDX, as visible in the plots. The Wilcoxon signed-rank test confirms this hypothesis.

\subsubsection{AA vs. DTX}

\begin{figure}[htbp]
    \centering
    \begin{subfigure}{0.494\columnwidth}
        \centering
        \includegraphics[width=1\linewidth]{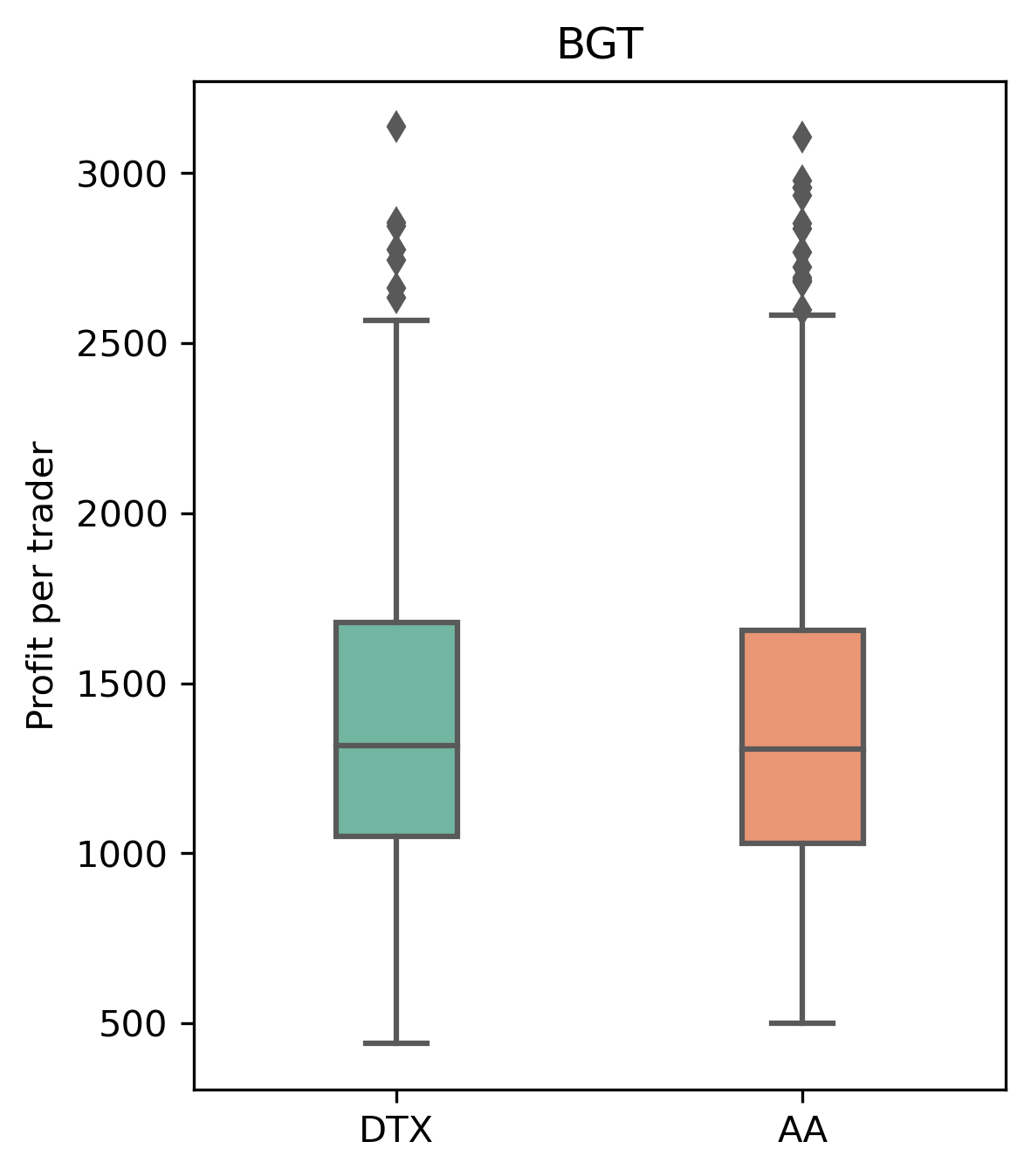}
        \caption{Balanced Group Tests for AA vs. DTX.}
        \label{fig:aa-bgr-dist}
    \end{subfigure}
    \begin{subfigure}{0.494\columnwidth}
        \centering
        \includegraphics[width=1\linewidth]{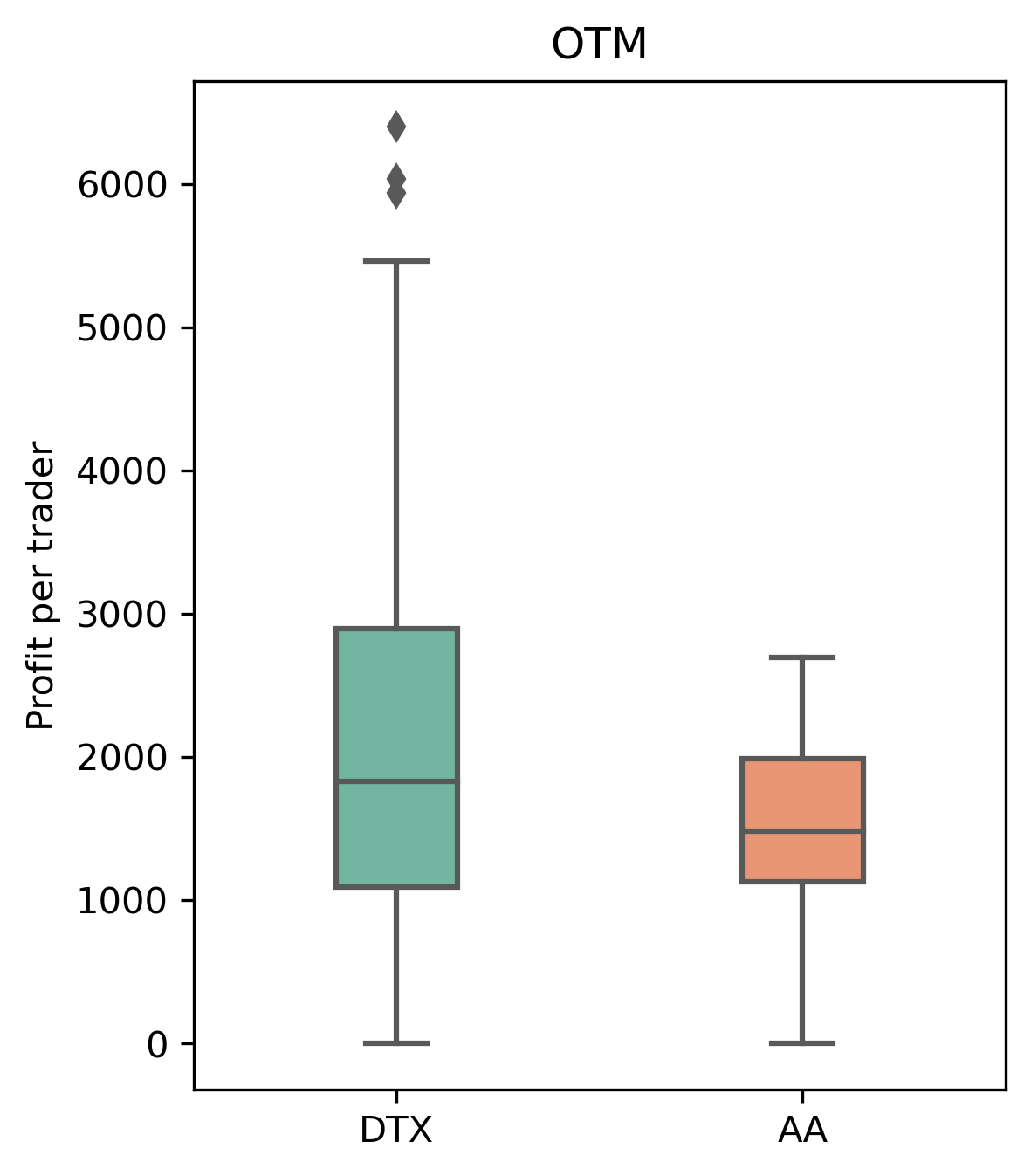}
        \caption{One to Many Tests for AA vs. DTX.}
        \label{fig:aa-otm-dist}
    \end{subfigure}
    \vspace{0.1cm}
    \caption{Box-plots showing PPT for ZIC vs. DTX tests.}
    \label{fig:aa-dist}
\end{figure}

\begin{figure}[!ht]
  \centering
   {\epsfig{file = 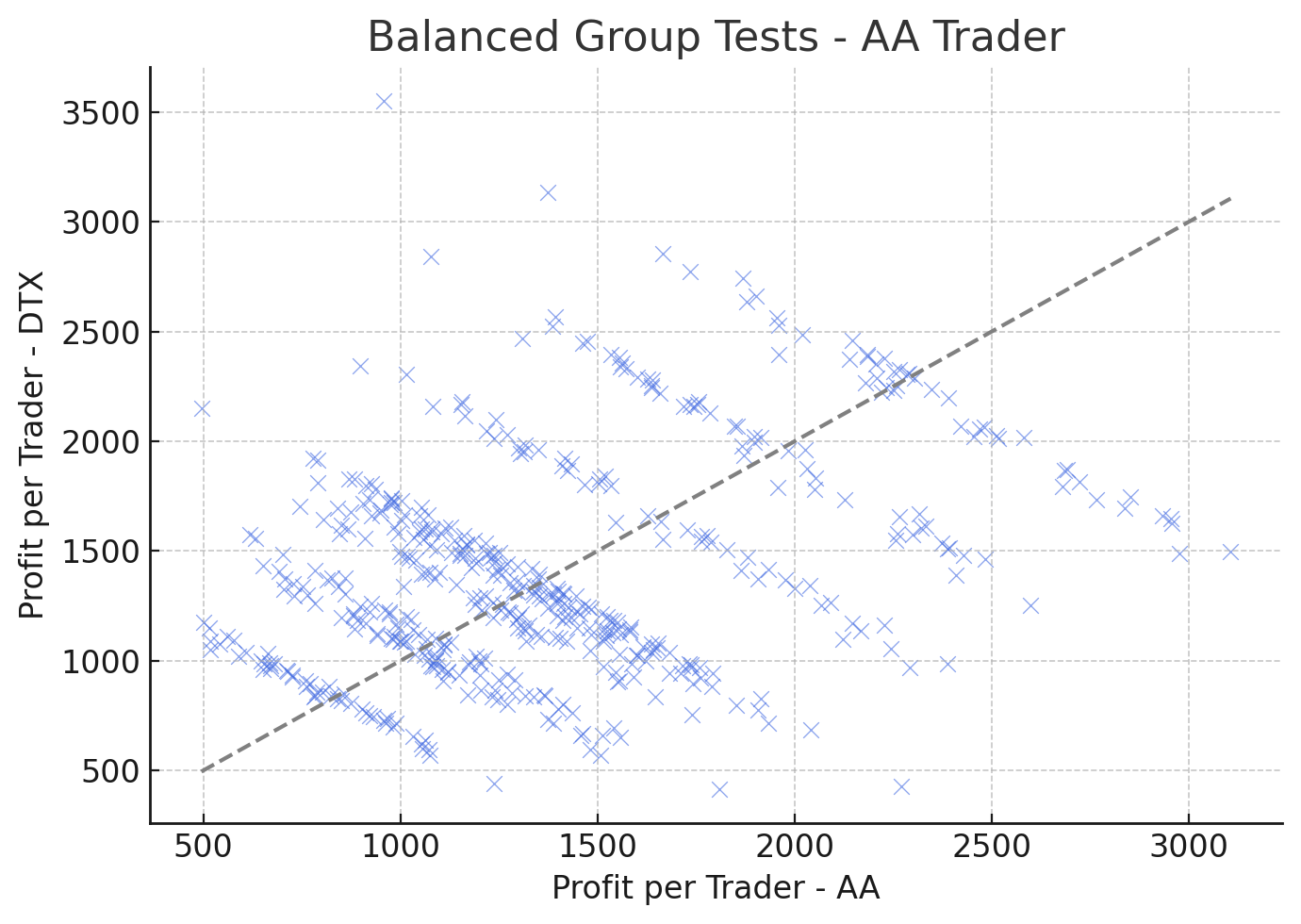, width = 7.5cm}}
  \caption{Scatter plot of PPT in BGTs for AA vs. DTX.}
  \label{fig:bgr-aa-dtx}
\end{figure}

Figures \ref{fig:aa-bgr-dist} and \ref{fig:bgr-aa-dtx} present a visual representation of the profits obtained by AA and DTX in the BGTs, indicating similar results for both traders. The statistical test applied failed to prove that there is a significant difference in terms of mean profits between AA and DTX. Thus, this is our only inconclusive experiment.

\begin{figure}[!ht]
  \centering
   {\epsfig{file = 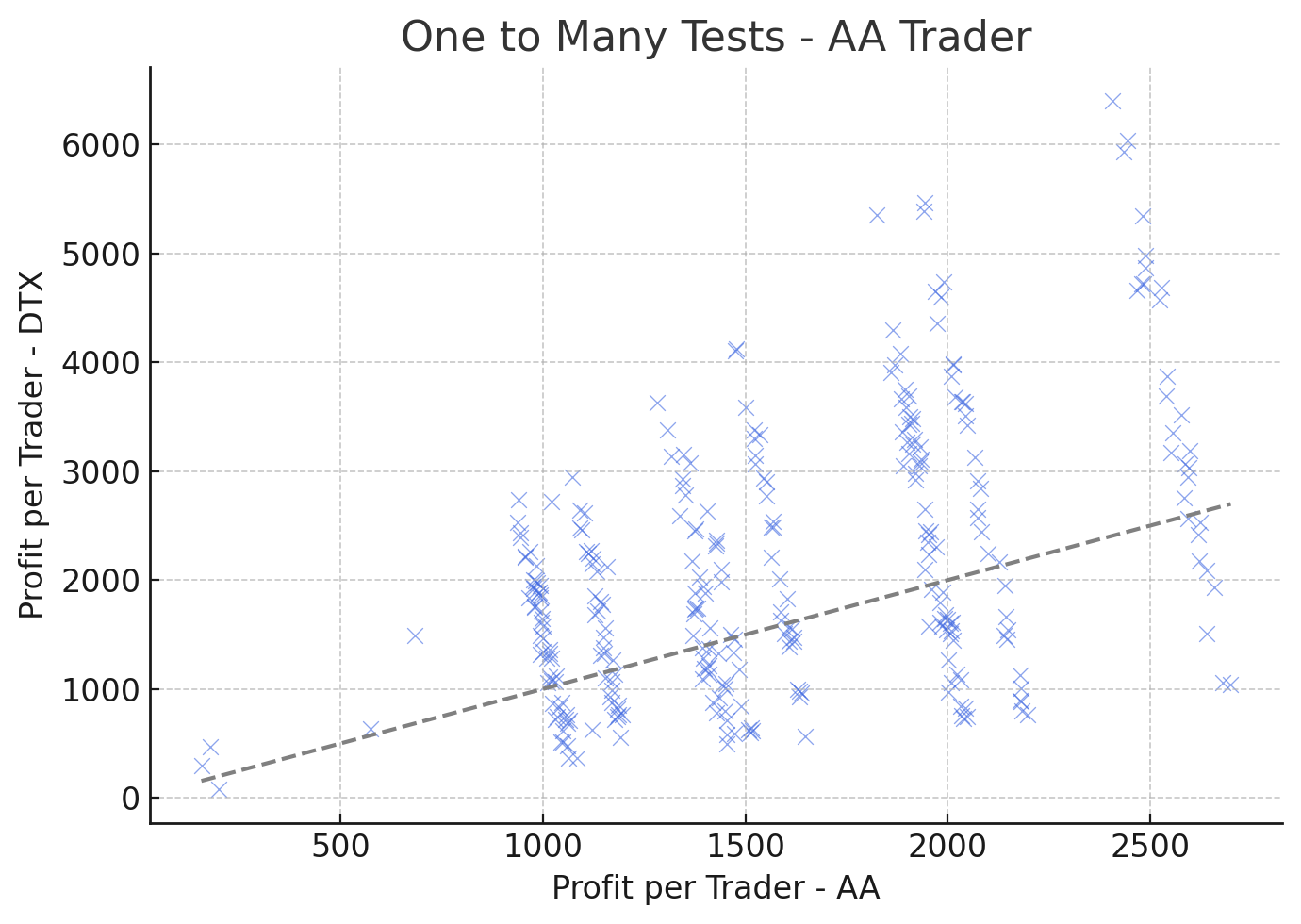, width = 7.5cm}}
  \caption{Scatter plot of PPT in OTMs for AA vs. DTX.}
  \label{fig:otm-aa-dtx}
\end{figure}

On the other side, Figure \ref{fig:aa-otm-dist} shows a high-profit but high-variance DTX in the OTM experiments against AA, a fact also visible by looking at the points above the diagonal in the scatter plot in Figure \ref{fig:otm-aa-dtx}. The statistical test concludes that DTX is the higher-performing strategy in this experiment, but with increased variance.

\subsubsection{Summary of Results}

The results presented in this section are used to objectively highlight what a trader based on a simple LSTM architecture is able to achieve. To recap, our model was exposed to market orders issued by traders and the corresponding LOB market data during training. The purpose of this is to enable the model to learn a mapping from the input, a variety of market statistics, and its own limit price, and generate an output, a quote that can be used to place an order. The success of a particular output quote is a combination of adaptability to the multi-threaded market's conditions, rapidity of response, and competitor's behaviour.

In summary, our empirical analyses reveal that DTX exhibits superior performance in six out of eight experiments and matches profits in one out of the eight. DTX achieves its sole purpose, which is to make profit. Notably, DTX either matches or surpasses the performance of three out of the four traders tested, including those deemed super-human. Specifically, DTX recorded two victories over GDX and exhibited a win-tie performance against AA. However, the results against Cliff's ZIP are more nuanced; DTX registered both a victory and a defeat in markets where both traded concurrently.

As we draw a line at the end of this section, having presented a detailed analysis of the results, we are nevertheless warranted to give credit to DTX, our Deep Learning powered trader, for its notable performance against the other traders in the literature. As we transition into the discussion, we will delve deeper into the implications of these results, discussing the strengths and weaknesses of DTX, relating its performance to previous results in TBSE, and exploring their broader impact on the field at the intersection of finance and artificial intelligence.

\section{\uppercase{Discussion}}
\label{section:discussion}

The central aim of our research was to create a Deep Learning model as part of a trading algorithm that trades in an asynchronous, more realistic market simulation, juxtaposing its efficacy against established trading strategies. Our findings, derived from rigorous experiments, offer valuable insights into the nuances of DTX's behaviour and its potential applications.

A salient observation from our results is the consistent profitability of DTX across all the experimental setups. This accomplishment is particularly noteworthy when considering the volatility of markets, which is better captured in the multi-threaded simulation that we use. DTX's ability to outperform or match traders such as AA, GDX, or ZIC suggests that the model can generalise effectively across various scenarios and aggregate on-the-spot information better than humans. Our trader relies on an opaque model that doesn't keep track of market changes, so it reacts consistently and quickly no matter the conditions.

However, its performance against traders like ZIP varied. Particularly, its profits have high variability, especially when trading as a defector. DTX does not behave under a pre-defined set of rules but instead tries to estimate the best price to trade at. Sometimes the model generates prices that would produce a loss if used, but DTX has a number of fallbacks that handle this behaviour, such as quoting prices right above or below the best bids or asks when that happens. These variations might be attributed to DTX not understanding the market but rather replicating pre-learned scenarios.

Within the broader academic discourse on trading algorithms, our findings resonate with \cite{Wray2020Automated}. They proposed this DLNN architecture, managing to outperform other strategies, but only trained it to copy specific traders in a sequential simulation.

When they introduced TBSE \cite{rollins2020trading}, Rollins and Cliff proposed the idea that the traders in the literature might have a different behaviour when tested in a concurrent simulation, better reflecting what real markets look like. The results they presented challenged the "status quo" of the trader dominance hierarchy, finding that they now come as follows: ZIP $>$ AA $>$ GDX $>$ ZIC. By quantifying the difference in results between DTX and the four traders, we can say that the relative performance of DTX follows the same ranking.

The broader implications of our research underscore the potential of Deep Learning trading algorithms in real-world trading scenarios. Their ability to be consistent, resilient, and generalise in any scenario suggests that they could be pivotal in creating fairer and more efficient markets. However, it's crucial to consider that markets populated solely by these intelligent automated systems might result in inexplicable events and our inability, as humans, to understand the new mechanisms of the financial markets we rely on.

\section{\uppercase{Limitations \& Future Work}}
\label{section:lim-futwork}

Our study, while comprehensive, is not without limitations. DTX was trained using rich data, but from only so many traders and scenarios. Also, our experimental setup was focused on only two types of traders at a time. Not to mention the considerable resource overhead involved in data collection, model training, and testing. Addressing these in future research would offer even more nuanced insights into DTX's capabilities. Moreover, an intriguing avenue for future exploration would be quantifying the correlation between the model's inference time and performance, as well as the degree of impact of each one of its 14 features.

In practical applications, a financial institution engaged in active trading could potentially deploy the DTX algorithm, provided they have access to extensive historical LOB data as well as their proprietary trading data. Given that access to limit order prices is typically restricted to an entity's own trading operations, DTX could be trained on this comprehensive dataset, thereby amalgamating the strengths of multiple established strategies and leveraging vast computing resources. A market populated with traders like DTX can be more efficient in allocating resources, creating a fair and predictable space. \cite{chatgpt}

\section{\uppercase{Conclusion}}
\label{section:conclussion}

In the rapidly evolving domain of automated trading, our study carves out a distinct space, emphasising the potential of Deep Learning trading algorithms. As markets continue to evolve, the quest for algorithms that can adapt and thrive remains paramount, and DTX, as evidenced by our research, stands as a promising proof-of-concept in this landscape.
In the quest for novelty and realism, we researched this in a distributed market simulation that has previously overturned the trader dominance hierarchy, with DTX being consistent with these findings.

As we stand on the cusp of this new frontier, it beckons researchers, practitioners, and policymakers alike to collaboratively shape a future where AI-augmented trading systems contribute to more efficient, stable, and equitable financial markets.

\section*{\uppercase{Acknowledgements}}

This paper would not have been possible without the guidance of my dissertation supervisor, Dave Cliff. There is also the need to mention that ChatGPT \cite{chatgpt} has helped with putting ideas into words, with a few parts of this paper written with its aid.

\bibliographystyle{apalike}
{\small
\bibliography{example}}

\end{document}